\documentclass[showpacs,preprintnumbers,superscriptaddress,amsmath,amssymb]{revtex4-2}
\usepackage[colorlinks=true,bookmarks=false,citecolor=blue,urlcolor=blue]{hyperref}
\usepackage{amsmath,amssymb}
\usepackage{graphicx}
\usepackage{verbatim}
\usepackage{enumitem}
\usepackage{color}
\usepackage[english]{babel}
\usepackage{bm}
\usepackage{natbib}
\usepackage[symbol]{footmisc}
\usepackage{textcomp}

\usepackage{epstopdf}
\usepackage{setspace}
\begin{document}
\title{Resonant binding of dielectric particles to metal surface without
plasmonics}

\author{Evgeny Bulgakov}
\affiliation{Kirensky Institute of Physics Federal Research Center
KSC SB RAS 660036 Krasnoyarsk Russia} \affiliation{Reshetnev
Siberian State University of Science and Technology, 660037,
Krasnoyarsk, Russia}
\author{Konstantin Pichugin}
\affiliation{Kirensky Institute of Physics Federal Research Center
KSC SB RAS 660036 Krasnoyarsk Russia}
\author{Almas Sadreev}
\affiliation{Kirensky Institute of Physics Federal Research Center
KSC SB RAS 660036 Krasnoyarsk Russia}
\date{\today}

\begin{abstract}
High index dielectric spherical particle supports the high-$Q$
resonant Mie modes that results in a regular series of sharp
resonances in the radiation pressure. A presence of perfectly
conducting metal surface transforms the Mie modes into the
extremely high-$Q$ magnetic bonding or electric anti-bonding modes
for close approaching of the sphere to the surface. We show that
the electromagnetic plane wave with normal incidence results in
repulsive or attractive resonant optical forces  relative to metal
for excitation of the electric bonding or magnetic anti-bonding
resonant modes respectively. A magnitude of resonant optical
forces reaches order of one nano Newton of magnitude for micron
size of silicon particles and power of light $1mW/\mu m^2$ that
exceeds the gravitational force by four orders. However what is
the most remarkable there are steady positions for the sphere
between pulling and pushing forces  that gives rise to resonant
binding of the sphere by metal surface. A frequency of mechanical
oscillations of particle around the equilibrium positions reaches
a magnitude of order MHz.
\end{abstract}
\maketitle
\section{Introduction}
An illumination of dielectric spherical particle by plane wave
with definite wave vector gives rise to radiation pressure which
drags the particle along the wave vector because of preservation
of momentum. Irvine has shown that regular series of sharp
resonances in the radiation pressure superimposes on this placid
background of radiation pressure caused by excitation of the Mie
resonant modes with high $Q$ factor in dielectric spherical
particles \cite{Irvine1965}. In 1977 Ashkin and Dziedzic reported
the first precise observation of these resonances based a force
spectroscopy \cite{Ashkin1977}. Illumination of  the spherical
particles by spatially structured beams can even pull particles
because of absence of definite longitudinal momentum
\cite{Novitsky2007,Chen2011,Li2020}. However irrespectively there
are no equilibrium positions for particles over the axis of beam.

Obviously, a presence of surface which reflects incident plane
wave can drastically change a situation with the equilibrium
positions for particles, i.e., the surface can trap particles
outside. Dielectric surface can trap particles using the
evanescent tails of waveguide modes \cite{Jaising2005} while metal
surface can trap particles due to excitation of surface plasmonic
waves by oblique plane waves
\cite{Volpe2006,Righini2007,Yang2011,Petrov2015,Ivinskaya2016,Maslov2017,Maslov2020}.
A key issue in particle trapping  by evanescent tails is the
formation of a stable equilibrium in the transverse direction
which is rather close to the surface while in the direction along
the surface the stability of particles is remaining in question.

In the present letter we show that a presence of perfectly
reflecting metal allows to avoid excitation of surface plasmonic
waves and to trap dielectric spheres at certain distances even for
normal incidence of plane wave. For the case of perfectly
conducting metal the problem of scattering  can be  solved  by
resorting  to image theory, according  to  which the field
scattered into the accessible half-space by a sphere in the
presence of the reflecting surface coincides with the field
scattered by the compound object  that includes  the  particle and
its  image, provided the exciting field is the superposition of
the actual incident field and  of the  field that  comes  from the
image source \cite{Borghese}. Therefore the problem can be
considered as a scattering of EM waves by two spheres. A presence
of the second particle or any scattering object leads to multiple
scattering between the object and particle and can lead to optical
binding of particle to the object even under illumination of one
beam. This is often referred to peculiar manifestation of optical
forces as optical binding, and   it was first discovered   by
Burns {\it et al} on a system of two plastic spheres   in water in
1989 \cite{Burns1989}. Depending on the particle separation, the
optical binding leads to attractive or repulsive forces between
the particles and, thus, contributes to the formation  of stable
configurations of particles.

Sharp features in the force spectroscopy, causing mutual
attraction or repulsion between successive photonic crystal layers
of dielectric spheres under illumination of plane wave has been
presented by Antonoyiannakis and Pendry
\cite{Antonoyiannakis1997}. Each layer is specified by extremely
narrow resonances which transform into the bonding and anti-boding
resonances for close approaching of the layers. It was revealed
that the lower frequency bonding resonance forces push the two
layers together and the higher frequency anti-bonding resonance
pull them apart. These conclusions are based on an analogy of the
Maxwell's equations with the quantum mechanics, in particular with
molecular orbitals. Later these disclosures we reported for
coupled photonic crystal slabs \cite{Liu2009}, two planar
dielectric photonic metamaterials \cite{Zhang2014},  two silicon
spheres \cite{Yano2017,Bulgakov2020a} and two coaxial disks
\cite{Bulgakov2020b}. Similarly for the case of dielectric sphere
with large refractive index at perfectly conducting surface the
high-$Q$ Mie resonant modes transform into the bonding TM type or
anti-bonding TE type modes which give rise to attractive or
repulsive resonant forces because of boundary conditions. And what
is important these forces considerably exceed by a two orders of
magnitude the resonant pushing forces in the case of bare sphere
\cite{Irvine1965,Ashkin1977}.

\section{Optical forces}
\begin{figure}
\centering \includegraphics*[width=9cm,clip=]{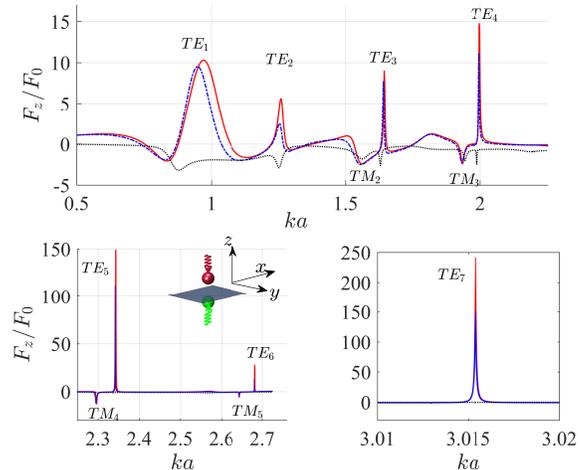} \caption{Force
normal to silver surface acting on silicon sphere with radius
$a=0.55\mu m$ and refractive index $n=3.464$ at wavelength
$\lambda=1.5 \mu m$  under illumination of plane wave with
intensity $P=1 mW/\mu m^2$. $F_0=8.1 pN$. Black dotted line shows
the radiation-pressure acting on bare sphere, red solid line shows
the case at distance $L/a=1.2$ from ideal metal, and blue
dash-dotted line shows the case of silver metal with refractive
index $n=0.14+11i$ at $\lambda=1.5\mu m$. Sub indices enumerate
orbital momentum $l$ of the Mie resonances of bare sphere. Inset
shows sphere (red) positioned at distance $L$ between center of
sphere and metal surface and its image (green). } \label{fig2}
\end{figure}

The problem is solved by T-matrix approach \cite{Mishchenko} that
allows us to calculate the optical force acting on the particle
via the stress tensor \cite{Antonoyiannakis1999}
\begin{eqnarray}\label{force}
&F_{\alpha}=\int T_{\alpha\beta}dS_{\beta}, &\nonumber\\
&T_{\alpha\beta}=
    \frac{1}{4\pi}E_{\alpha}E_{\beta}^{*}-\frac{1}{8\pi}\delta_{\alpha\beta}
|{\bf
E}|^2+\frac{1}{4\pi}H_{\alpha}H_{\beta}^{*}-\frac{1}{8\pi}\delta_{\alpha\beta}
   |{\bf H}|^2.&
\end{eqnarray}
In what follows we consider plane wave incident normally to the
metal surface as shown by red wavy line in inset of Fig.
\ref{fig2}. That makes the system  equivalent to two identical
spheres in air in the superposed standing wave
$E_0\overrightarrow{e}_x\sin kz$ with even solution for $E_z$ and
odd for $H_z$. Respectively we have odd/even solutions for the
tangential components of electric/magnetic fields. The results of
calculations are presented in Fig. \ref{fig2} where
dotted line shows the radiation-pressure force onto the bare
sphere in air with peaks located at the Mie resonant modes indexed
by orbital momentum $l$.
That approach differs from consideration of optical forces
effected dielectric particles by metal surfaces due to excitation
surface plasmon modes by oblique plane waves
\cite{Righini2007,Yang2011,Petrov2015,Maslov2018}.

\begin{figure}
\centering \includegraphics*[width=9cm,clip=]{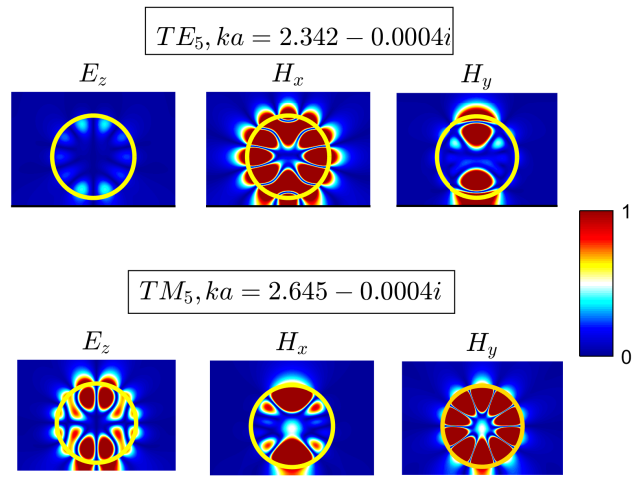}
\caption{Absolute values of components of EM field of resonant
modes at the distance $L=1.2a$ from ideally conducting surface
with corresponding complex resonant frequencies.} \label{fig3}
\end{figure}

When two spheres in air are approaching the Mie resonances of each
sphere are split and their resonant modes are hybridized forming
the bonding and anti-bonding resonant modes irrespective to
polarization. Respectively, a sign of optical force between two
spheres follows these modes \cite{Bulgakov2020a}. For the present
case of one sphere near metal surface the Mie resonances are not
split for approaching of sphere to surface but only shifted as
force spectroscopy shows in Fig. \ref{fig2}. The reason is related
to that the odd standing wave field $\sin kz$ effecting the sphere
and its image can excite only odd resonant modes for tangential
components of electric field.
 One can see that for the TE resonances the shift is
positive with repulsive optical forces while for the TM resonances
the shift is negative with attractive forces. In order to
comprehend this rule we show in Fig. \ref{fig3} the resonant modes
$TE_5$ and $TM_5$ from the perfectly conducting surface.
For the $TE_5$ one can see the transverse components of magnetic
field dominate compared to the $E_z$ in space between sphere and
surface. Therefore the stress-tensor component can be approximated
as $T_{zz}\approx -\frac{1}{8\pi}(|H_x|^2+|H_y|^2)<0$ to result in
positive, i.e., the repulsive force that indeed fully agrees with
Fig. \ref{fig2} for TE resonances. As for as the resonant mode
$TM_5$ one can see from Fig. \ref{fig3} that $E_z$ dominates over
the tangential magnetic field to have $T_{zz}\approx
\frac{1}{8\pi}(|E_z|^2-|H_x|^2-|H_y|^2)$ positive. Therefore the
TM resonances result in attractive force as also agrees with Fig.
\ref{fig2}. These results are hold also for the case of real
silver metal surface with the refractive index $n=0.14+11i$ with
however slightly reduced values of the optical force as shown in
Fig. \ref{fig2} by blue dash-dotted line.

\begin{figure}
\centering \includegraphics*[width=9cm,clip=]{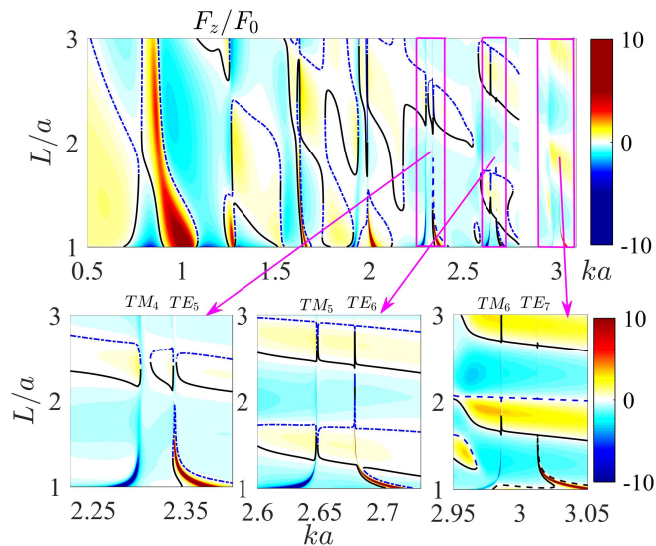} \caption{The
optical force  between sphere and ideally conducting surface vs
the frequency and distance for illumination of plane mode with
normal incidence. Blue dash solid  lines show steady positions  of
sphere while black solid lines show unsteady positions.
$F_0=8.1pN$.} \label{fig4}
\end{figure}

In Fig. \ref{fig4} we show results of calculations of optical
force versus the frequency of plane wave with normal incidence and
power $1 mW/\mu m^2$ and distance between sphere and metal
surface. Because of oscillating behavior of the optical force
\cite{Karasek2006,Bulgakov2020a} one can see that attractive
resonant forces around the $TM_{l-1}$ resonances is substituted by
repulsive forces around the $TE_l$ resonances.
Similar to Fig. \ref{fig2} Fig. \ref{fig4} manifests force
spectroscopy by  Ashkin and Dziedzic \cite{Ashkin1977}. Fig.
\ref{fig4} shows also equilibrium positions which depend on
distance $L$. What is remarkable there are areas with equilibria
of the sphere almost independent of $L$ for high indices $l$.
Therefore is to assume that silicon particles of different sizes
flow normally to metal surface the particles with definite sizes
will bind at definite distances realizing selection by size. That
brings new potential to exploit these results for optical trapping
and for optical resonant sorting \cite{Shilkin2017}.

\begin{figure}
\centering\includegraphics*[width=7cm,clip=]{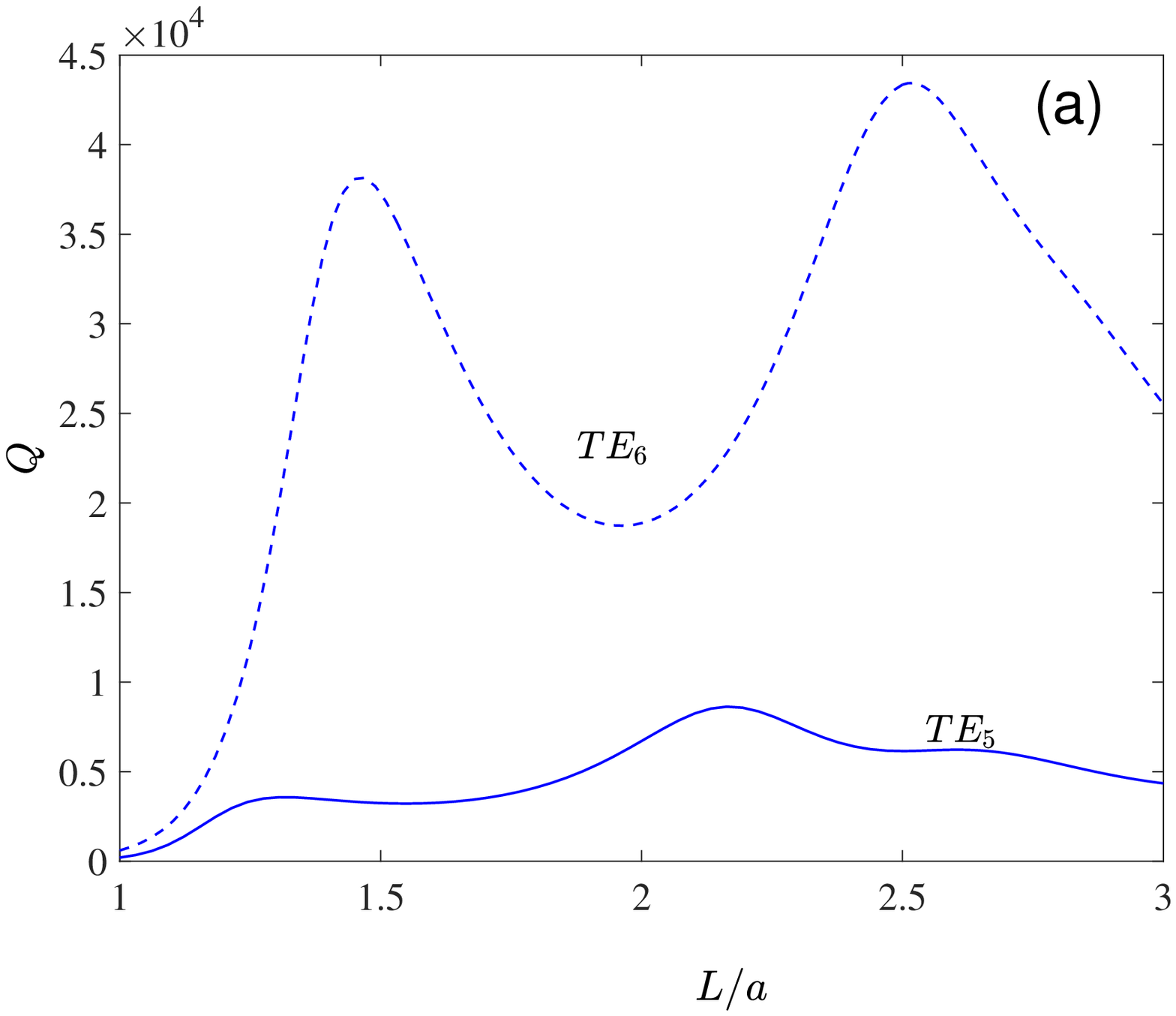}
\includegraphics*[width=7cm,clip=]{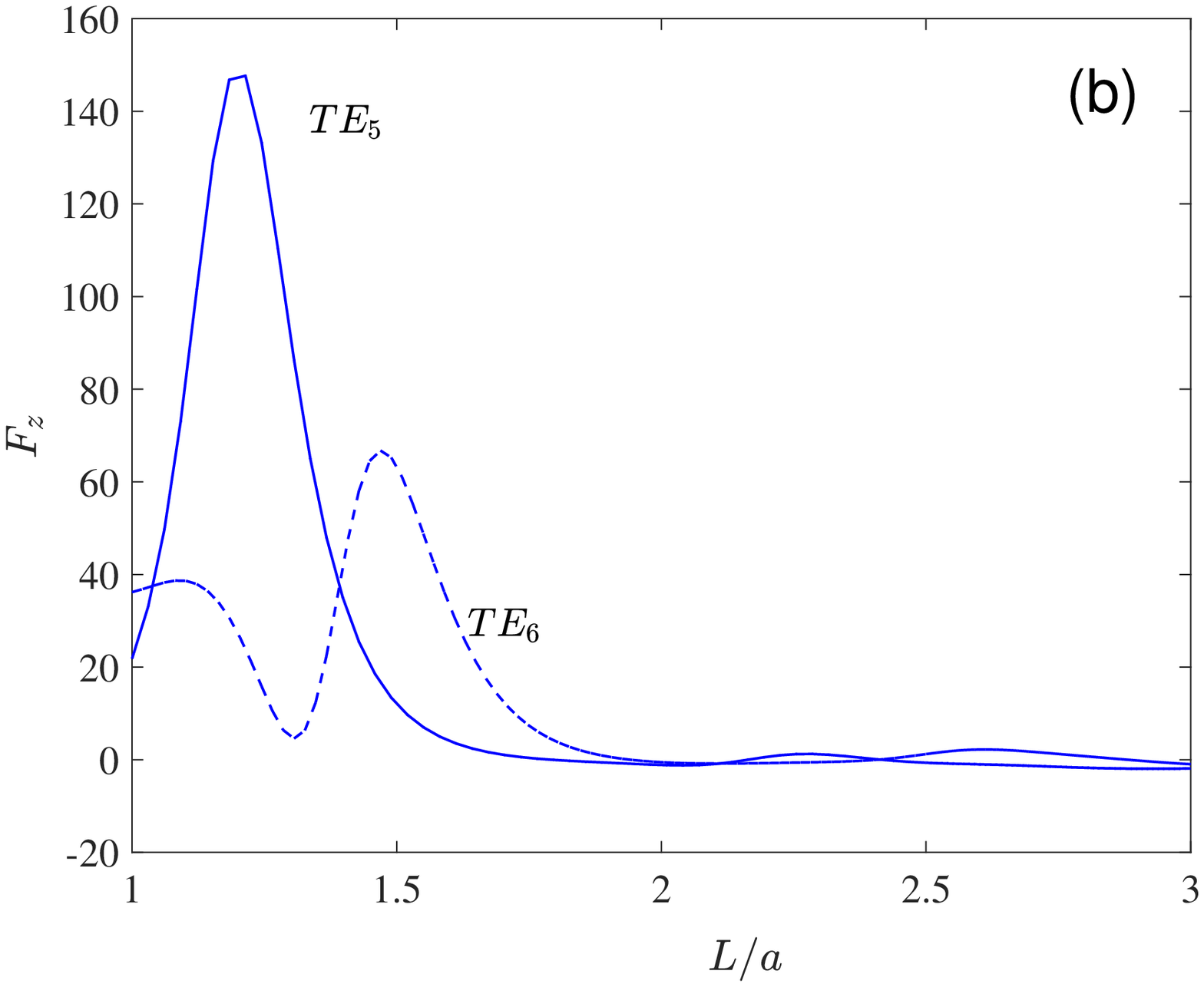} \caption{The $Q$ factor
(a) and resonant force (b) for the case of excitation of the
$TE_5$ and $TE_6$ resonant modes by plane wave with power $1mW/\mu
m^2$ vs the distance between particle and metal silver surface.}
\label{fig5}
\end{figure}

The insets in Fig. \ref{fig4} demonstrate considerable enhancement
of optical force up to hundreds of pico Newtons that exceed the
gravitational forces by four orders for micron size silicon
particles. Expectedly, such an enormous enhancement of resonant
forces is the result of extremely large $Q$-factors of the
resonances as was considered in papers
\cite{Antonoyiannakis1997,Liu2017}. However Fig. \ref{fig5} shows
that for the present case there is no a strict correlation between
the $Q$ factor of resonant mode and the resonant force. Although
the $Q$ factor as well as the forces grow with increasing of
orbital momentum $l$ as Fig. \ref{fig2} shows nevertheless we see
that the resonant force $TE_5$ exceeds the resonant force $TE_6$.
The reason is related to that not only the $Q$ factor contributes
to the resonant optical force but also the coupling constant of
incident field with the corresponding resonant mode. In
particular, Liu {\it et al} have observed already that the
higher-Q state exhibits a much weaker coupling to the external
radiation \cite{Liu2017}.

\section{Conclusions}
\begin{figure}
\centering \includegraphics*[width=9cm,clip=]{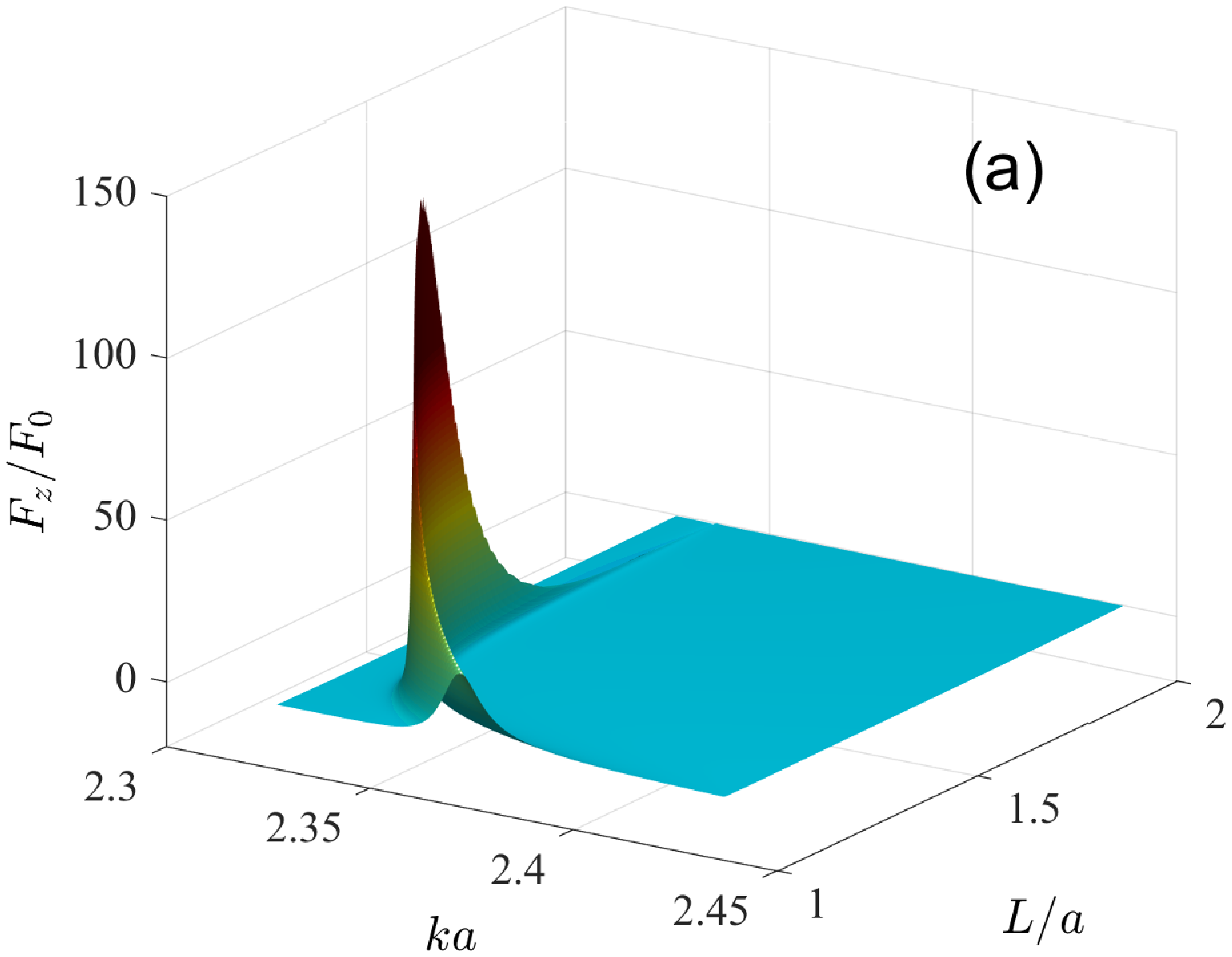}
\includegraphics*[width=8cm,clip=]{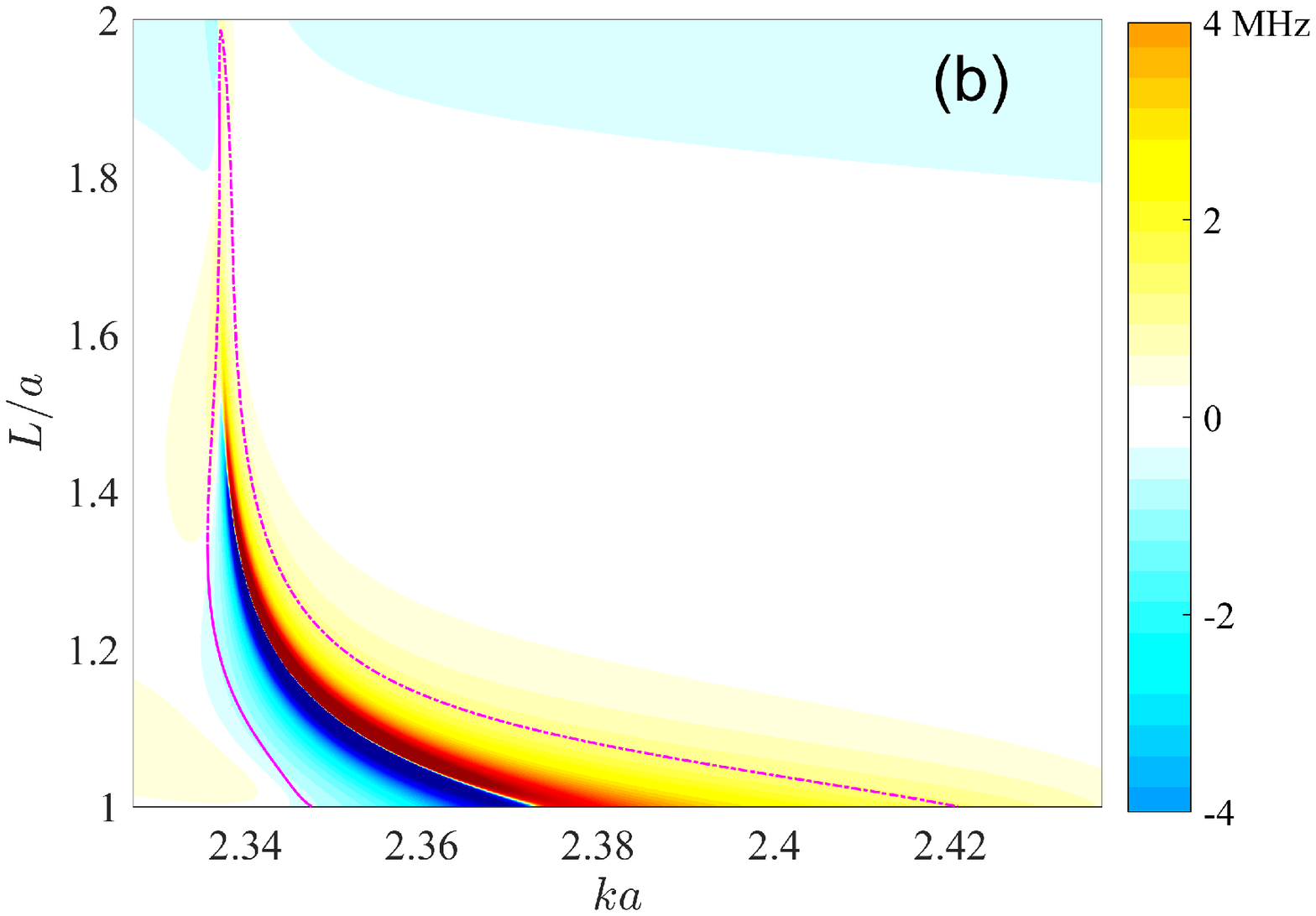}
\caption{Optical force (a) and respectively its derivative
$\sqrt{\frac{1}{m}|\frac{\partial F_z}{\partial z}|}$ (b) of
silicon particle with a mass $m$ around steady positions shown by
dash-doted (stable) and solid (unstable) lines.} \label{fig6}
\end{figure}

One can expect that the larger optical forces will take away the
sphere into positions far from the resonance in order to diminish
the forces. Indeed, this conclusion holds for the resonances with
lower $l$. However, for the extremely high $Q$ resonances with
$l=5,6,...$ one can see that the steady distances of silicon
sphere from the metal surface are almost independent of $L$ and
very close to the resonant frequencies. That tendency for
particles to place into positions close to resonances was revealed
also in Fabry-Perot resonator with movable mirrors
\cite{Sadreev2016}. By differentiation of the force  over $z$ it
is easy to find vibrational frequency of sphere around equilibrium
positions.
Fig. \ref{fig6} shows that the vibrational frequency is around of
a few MHz near steady positions of sphere except a vicinity of
resonance $TE_5$ where the frequency enhances by one order in
magnitude.

Therefore we can conclude that metal surface can strongly bind
silicon spherical particles under normal incidence of plane wave
with power $1 mW/\mu m^2$. However there are two regimes of
binding. In the first regime particles are desorbed by metal
surface by extremely large optical forces shown in the insets of
Fig. \ref{fig4} by deep blue color. And this process depends on
the sizes of particles. In the second regime the particles
levitates at the definite distances from the metal surface owing
to steady positions shown by blue dash lines in Fig. \ref{fig4}.
One can see that these distances of resonant levitation are
sensitive to the radius of spheres. Therefore if to assume that
the spheres of different sizes are flowing in viscous liquid the
plane wave mostly will desorb spheres on surface of metal. However
some fracture of particles of definite sizes will be trapped on
definite distances from the surface of meta that can pave a way
for resonant sorting of particles by size.\\ \ \\

{\bf Acknowledgements}\\
 The work was supported by
Russian Foundation for Basic Research projects No. 19-02-00055 and
Russian Scientific Foundation No. 21-12-00131.

\bibliography{sadreev}
\end{document}